\documentclass[12pt,a4paper]{scrartcl}
\usepackage{amsmath}
\usepackage{amssymb}
\usepackage[english]{babel}
\usepackage{bbm}
\usepackage{parskip}
\usepackage{setspace}
\usepackage{slashed}

\newcommand{\sign}{\text{sign}}

\newcommand{\ud}{\text{d}}
\newcommand{\ui}{\text{i}}
\newcommand{\ue}{\text{e}}

\newcommand{\di}{\slashed{\partial}}

\newcommand{\pslsh}{\slashed{\partial}}
\newcommand{\mcB}{\mathcal{B}}
\newcommand{\id}{\mathbbm{1}}
\newcommand{\reals}{\mathbbm{R}}
\newcommand{\nats}{\mathbbm{N}}

\newcommand{\ket}[1]{\mathinner{|{#1}\rangle}}
\newcommand{\vac}{\ket{\Omega}}
\newcommand{\p}{\partial}
\newcommand{\h}{\frac{1}{2}}
\newcommand{\refs}[1]{(\ref{#1})}
\newcommand{\beq}{\begin{eqnarray}}
\newcommand{\eeq}{\end{eqnarray}}
\newcommand{\nn}{\nonumber}
\newcommand{\Th}{\theta}

\newtheorem{theorem}{Theorem}[section]
\newtheorem{lemma}[theorem]{Lemma}
\newtheorem{remark}[theorem]{Remark}

\def\Tr{\operatorname{Tr}}
\def\dfi{d_5 (\theta , D)}
\def\dsi{d_6 (\theta , D)}

\def\dtw{d_2 (\theta , D)}
\def\dth{d_3 (\theta , D)}
\def\dfo{d_4 (\theta , D)}
\def\don{d_1 (\theta , D)}
\def\Th{\theta}
\def\pip{\Pi_+}

\def\pipl{\Pi_+^\star}

\def\gt{\gamma_*}

\def\cal{\mathcal}

\begin{document}
\title{Spectral asymmetry on the ball and asymptotics of the asymmetry kernel}
\author{A. Kirchberg\\
Theoretisch-Physikalisches Institut,
Friedrich-Schiller-Universit{\"a}t Jena\\
Max-Wien-Platz 1, 07743 Jena, Germany\\[.2cm]
K. Kirsten\thanks{e-mail: Klaus\_Kirsten@baylor.edu}\\
Department of Mathematics, Baylor University\\ Waco, TX 76798,
USA\\[.2cm]
E. M. Santangelo\thanks{e-mail: mariel@obelix.fisica.unlp.edu.ar}\\
Departamento de F\'{\i}sica, Universidad Nacional de La Plata\\
C.C.67, 1900 La Plata, Argentina\\[.2cm]
A. Wipf\thanks{e-mail: wipf@tpi.uni-jena.de}\\
Theoretisch-Physikalisches Institut,
Friedrich-Schiller-Universit{\"a}t Jena\\
Max-Wien-Platz 1, 07743 Jena, Germany}
\maketitle
\begin{abstract}
Let $\ui\di$ be the Dirac operator on a $D=2d$ dimensional ball
$\mcB$ with radius $R$. We calculate the spectral asymmetry
$\eta(0,\ui\di )$ for $D=2$ and $D=4$, when local chiral bag
boundary conditions are imposed. With these boundary conditions,
we also analyze the small-$t$ asymptotics of the heat trace $\Tr
(F P e^{-t P^2})$ where $P$ is an operator of Dirac type and $F$
is an auxiliary smooth smearing function.
\end{abstract}
\setcounter{tocdepth}{1}
\section{Introduction}

Local boundary conditions of chiral bag type for Euclidean Dirac operators have attracted,
over many years, the interest of both mathematicians and physicists.

From the physical point of view, such conditions are closely
related to those appearing in the effective models of quark
confinement known as chiral bag models \cite{goldstone}. They
contain a real parameter $\theta$, which is to be interpreted as
the analytic continuation of the well known $\theta$-parameter in
gauge theories. Indeed, for $\theta\neq 0$, the effective actions
for Dirac fermions contain a $CP$-breaking term proportional to
$\theta$ and proportional to the instanton number
\cite{balog,Wipf:1995dy}. Moreover, these boundary conditions can
be applied to one-loop quantum cosmology \cite{death,
esposito},supergravity theories \cite{deser} and branes
\cite{lugo} and are important in the investigation of conformal
anomalies \cite{maravassil}. Applications to finite temperature
problems were studied in \cite{Wipf:1995dy,d-w,bene-04}

From a mathematical point of view, as shown in \cite{Wipf:1995dy},
Euclidean Dirac operators under local boundary conditions of
chiral type define self-adjoint boundary problems. More recently,
it was shown \cite{bene03-36-11533} that both the first order
boundary value problem and its associated second order one are
strongly elliptic. The asymptotics of the smeared trace of the
heat kernel was studied, for the second order problem on general
Riemannian manifolds with boundary, in \cite{E-G-K-05}, where use
was made of functorial methods \cite{gilk95b,kkbuch,vassil} and
special case calculations presented in
\cite{E-K-02,bene03-36-11533}; in this context see also
\cite{G-K-05}.

The main characteristic of these boundary conditions, as compared
to nonlocal or Atiyah-Patodi-Singer \cite{APS} ones, is the
explicit breaking of chiral symmetry. Since the Dirac operator has
no zero modes this breaking comes only from the excited modes.
Since the asymmetry is encoded in the corresponding eta-function,
the study of this function if of central importance, physical
applications for example arising in the analysis of fermion number
fractionization in different field theory models
\cite{new12,new13,new17,new14}.

A recent step forward in the analysis of the eta function is
\cite{G-K-05}, where the eta function associated to the boundary
value problem was shown to be regular at $s=0$, and where some
properties of the asymptotic coefficients in the trace of the
smeared kernel corresponding to the eta function were obtained.

In the present paper, we evaluate the spectral asymmetry for Dirac
operators on the ball in two and four Euclidean dimensions, with
their domain defined by local boundary conditions of chiral bag
type. Moreover, making use of these results, and the corresponding
ones for cylindrical manifolds obtained in
\cite{bene03-36-11533,BSW} we determine some properties of the
leading asymptotic coefficients in the trace of the smeared kernel
corresponding to the eta function, for arbitrary Riemannian
manifolds.

The outlay of the paper is as follows:  In section
\ref{eigenvalue} we present a new method, based on Group Theory,
to obtain the spectrum of the boundary value problem under study.
Section \ref{eta} presents an integral formula for the spectral
asymmetry of the Euclidean Dirac operator in a ball of arbitrary
even dimension. Section \ref{section3} is devoted to the explicit
calculation of the asymmetry in a disk, while the specral
asymmetry in the four dimensional case is evaluated in section
\ref{four}. Finally, in section \ref{coefficients}, some
properties of the leading coefficients in the heat trace
associated to the eta function are established through functorial
methods, and the use of special cases. To be self-contained, the
Appendix contains a derivation of the Debye expansion of the
Bessel functions.

\section{Eigenvalue Problem for the Dirac Operator}
\label{eigenvalue} Although the eigenvalue equations for the
problem at hand were derived before \cite{DK,E-K-02}, we present
here a different derivation, based on group theory. We write the
free Euclidean Dirac operator (i.e. gauge fields are absent) as
\begin{align}
\label{eig1} \ui \di &
  = \ui \sum_{A=1}^{D} \gamma^A \frac{\p}{\p x_A}
  = \ui \gamma^A \partial_A,
\end{align}
where we used the Einstein summation convention for the index
$A=1,\ldots,D,$ and the Dirac-matrices fulfill the Clifford algebra
\begin{align}
\label{eig2}
\{ \gamma_A , \gamma_B \} &= 2 \delta_{AB}.
\end{align}
We have to choose boundary conditions for the spinors $\Psi$, such
that the Dirac operator is Hermitian and in the following we will
consider chiral-bag boundary conditions. Given the projection
operators
\begin{align}
\label{eig3} \Pi_\mp &= \h(\id \mp \ui \gamma_* \ue^{\gamma_*
\theta} S)
\end{align}
with free parameter $\theta$ and
\begin{align}
\label{eig4}
\gamma_* &= (-\ui)^d \gamma_1 \cdots \gamma_D, \quad
S = \gamma^A x_A /r, \quad
r = \sqrt{x^A x_A},
\end{align}
these boundary conditions are defined as
\begin{align}
\label{eig5} \Pi_- \Psi |_{\p\mcB} &= 0.
\end{align}
In particular, $S$ is the projection of $\gamma^A$ onto the
\emph{outward} unit normal vector. In contrast to
Atiyah-Patodi-Singer boundary conditions, the chiral-bag boundary
conditions are local. One can further show that, for simply
connected boundaries, these boundary conditions do not allow for
zero modes. Further details can be found in \cite{Wipf:1995dy}.

$P$ and $\Pi_-$ commute with the total angular momentum
\begin{align}
\label{eig6}
J_{AB} & = L_{AB} + \Sigma_{AB}, \quad
L_{AB} = - \ui (x_A \p_B - x_B \p_A), \quad
\Sigma_{AB} = \frac{1}{4 \ui} [ \gamma_A , \gamma_B ].
\end{align}
Using the algebraic approach developed in \cite{Kirchberg:2002me},
we can first diagonalize the total angular momentum, i.e.
determine the \emph{spin spherical harmonics} by group theoretical
methods. Our aim is to construct the highest weight states. These
states are eigenstates of the Cartan operators of the
$\mathfrak{so}(D)$-algebra and are annihilated by the
corresponding raising operators. The remaining states in each
multiplet can be obtained by applying lowering operators on these
highest weight states. It is appropriate to introduce complex
coordinates and creation/annihilation operators as follows,
\begin{align}
\label{eig7} z_a & = x_{2a-1} + \ui x_{2a}, \quad
\partial_a = \h (\p_{x_{2a-1}}- \ui \p_{x_{2a}}), \quad \\
\nn \psi^\dagger_a & = \h ( \gamma_{2a-1} + \ui \gamma_{2a}),
\quad \psi_a = \h ( \gamma_{2a-1} - \ui \gamma_{2a}), \quad a =
1,\ldots,d =D/2.
\end{align}

For the fermionic operators, one easily verifies the relations
\begin{align}
\label{eig8}
\{\psi_a, \psi^\dagger_b \} = \delta_{ab}, \quad
\{\psi_a, \psi_b \} = 0, \quad
\{\psi_a^\dagger, \psi_b^\dagger \} = 0.
\end{align}
Thus, $\psi^\dagger_a$ ($\psi_a$) acts as a creation (annihilation)
operator, and we can employ the usual (fermionic) Fock space construction,
starting with the vaccum state $\vac$,
\begin{align}
\label{eig9}
\psi_a \vac = 0, \quad
\ket{a_1 \ldots a_m} \equiv \psi^\dagger_{a_1} \ldots \psi^\dagger_{a_m} \vac, \quad
1 \le m \le d.
\end{align}
Now we choose the following Cartan-Weyl basis for the
$\mathfrak{so}(D)$-algebra: Cartan operators $H_a$ and raising
operators $E_1,...,E_d$ (corresponding to simple positive roots)
read (no sum!)
\begin{align}
\label{eig10}
\nn
H_a & = z_a \partial_a - \bar{z}_a \bar{\partial}_a + \h
( \psi_a^\dagger \psi_a - \psi_a \psi_a^\dagger ), \quad a=1,\ldots,d,\\
E_a & = - \ui ( z_a \partial_{a+1} - \bar{z}_{a+1} \bar{\partial}_a +
\psi_a^\dagger \psi_{a+1}),  \quad a=1,\ldots,d-1, \\
\nn
E_d & = - \ui ( z_{d-1} \bar{\partial}_d - z_d \bar{\partial}_{d-1}
+ \psi^\dagger_{d-1} \psi^\dagger_d ).
\end{align}
Since the operators in \refs{eig10} act trivially on the radial
part of spinor wave functions, we consider the angular part only.
We can easily determine the highest weight states with respect to the orbital part,
they are given by
\begin{align}
\label{eig12} \phi_\ell = z_1^\ell .
\end{align}
These states are annihilated by all simple positive roots, and the
eigenvalues with respect to the Cartans read $(H_1, \ldots,
H_d)=(\ell,0,\ldots,0)$. Similarly, for the `fermionic' part there
are only two highest weight states given by
\begin{align}
\label{eig13}
\chi^+= \psi_1^\dagger \cdots \psi_d^\dagger \vac \quad
\text{and} \quad \chi^- = \psi_1^\dagger \cdots
\psi_{d-1}^\dagger \vac.
\end{align}
The corresponding eigenvalues of the Cartan operators read $(\h,
\ldots, , \h, \pm \h)$, respectively. Next, we determine the
highest weight states of `fermionic' and `bosonic' degrees of
freedom together. Two highest weight states can be constructed
easily; they are just given by the tensor products
\begin{align}
\label{eig14} \phi_\ell^+ = \phi_\ell \chi^+ \quad \text{and}
\quad \phi_\ell^- = \phi_\ell \chi^-,
\end{align}
with eigenvalues of the Cartan operators
$(\ell+\h,\h,\ldots,\h,\pm\h)$, respectively. Furthermore, we
observe that the operator $S$ in \refs{eig4},
\begin{align}
\label{eig15} S = (\psi_a^\dagger \bar{z}_a + \psi_a z_a) /r =
S^\dagger, \qquad S^2 = \mathbbm{1},
\end{align}
commutes with the total angular momentum and therefore maps highest weight
states into highest weight states.
We obtain two additional highest weight states,
\begin{align}
\label{eig16} \tilde{\phi}_\ell^+ = S \phi_\ell^+ \quad \text{and}
\quad \tilde{\phi}_\ell^- = S \phi_\ell^-,
\end{align}
with eigenvalues $(\ell+\h,\h,\ldots,\h,\pm\h)$. Equations
\refs{eig14} and \refs{eig16} contain, in fact, all highest weight
states, which is seen as follows. The tensor product of the
bosonic representation $(\ell ,0,\ldots,0)$ with the fermionic
representation $(\h, \ldots,\h)$ is given by
\begin{align}
\renewcommand{\arraystretch}{1.5}
\begin{array}{ccccccc}
(\ell,0,\ldots,0) & \otimes & (\h, \ldots,\h) & = &
(\ell+\h,\h,\ldots,\h) &
\oplus & (\ell-\h,\h,\ldots,\h,-\h) \\
\phi_\ell && \chi^+  && \phi_\ell^+ && \tilde{\phi}_{\ell-1}^-
\end{array}
\end{align}
and the tensor product of the bosonic representation
$(\ell,0,\ldots,0)$ with the fermionic representation $(\h,
\ldots,\h,-\h)$ is given by
\begin{align}
\renewcommand{\arraystretch}{1.5}
\begin{array}{ccccccc}
(\ell,0,\ldots,0) & \otimes & (\h, \ldots,\h,-\h) & = &
(\ell+\h,\h,\ldots,\h,-\h) &
\oplus & (\ell-\h,\h,\ldots,\h) \\
\phi_\ell && \chi^- && \phi^-_\ell && \tilde{\phi}_{\ell-1}^+.
\end{array}
\end{align}
The highest weight states appear at the correct places, which is
determined by the degree of the polynomials in $x_A$ and the
chirality of the states. By counting the dimensions of the
representations and comparing it with Weyl's dimension formula for
the $\text{D}_d$ groups \cite{Kirchberg:2002me}, we see that there
are no further representations in the tensor product rule.

Next, we investigate the chiral-bag boundary conditions. As stated
above, $\Pi_-$ commutes with $J_{AB}$, and we can diagonalize
$\Pi_-$ in our basis of highest weigth states. We may express
$\gamma_*$ with the help of \refs{eig7} by
\begin{align}
\label{eig17}
\gamma_* = \prod_{a=1}^d (\psi_a^\dagger \psi_a - \psi_a \psi_a^\dagger),
\quad \text{such that} \quad \gamma_* \ket{\Omega} = (-)^d \ket{\Omega}.
\end{align}
Since $\phi_\ell^+$ and $\tilde{\phi}^+_\ell$ (and likewise
$\phi_\ell^-$ and $\tilde{\phi}_\ell^-$) have the same eigenvalues
with respect to the Cartan operators $H_a$, we allow for linear
combinations of them,
\begin{align}
\label{eig18} \Psi^\pm_\ell &= f^\pm_\ell (r) \phi^\pm_\ell +
g_\ell^\pm(r) \tilde{\phi}^\pm_\ell.
\end{align}
Imposing the boundary condition $\Pi_- \Psi^\pm_\ell=0$ leads to
the following equations for the radial functions
\begin{align}
\label{eig19} f_\ell^\pm(R) \mp \ui \ue^{\pm \theta} g_\ell^\pm(R)
= 0.
\end{align}
Finally, using complex coordinates, we solve for the spectrum of
the Dirac operator,
\begin{align}
\label{eig20} \ui \pslsh \Psi^\pm_\ell & = \lambda \Psi^\pm_\ell,
\quad \ui \pslsh = 2 \ui \psi_a \bar{\p}_a + 2 \ui \psi^\dagger_a
\p_a.
\end{align}
From the eigenvalue equation of the Dirac operator \refs{eig20},
one obtains a system of coupled first-order differential
equations. Using $\nu = \ell + d -1/2$, the system reads \beq \ui
\left( f_\ell^\pm\right)' = \lambda g_\ell^\pm ,\quad \ui \left(
g_\ell ^\pm\right)' = \lambda f_\ell^\pm - \frac {2\ui} r \,\, \nu
g_\ell^\pm .\nn\eeq It can be easily solved, and the solutions are
given by
\begin{align}
\label{eig24} f^\pm_\ell(r) &= r^{1/2-\nu} (c_1 J_{\nu -1/2
}(|\lambda|r) + c_2
N_{\nu -1/2}(|\lambda|r)), \\
\label{eig25} g^\pm_\ell(r) &= - \ui \,\,\sign(\lambda)
r^{1/2-\nu} (c_1 J_{\nu + 1/2}(|\lambda|r) + c_2 N_{\nu
+1/2}(|\lambda|r)).
\end{align}
Finally, keeping only those eigenfunctions that are nonsingular at the origin,
imposing the boundary condition (\ref{eig19}) and defining
$k \equiv |\lambda| R$, we obtain
\begin{align}
\label{eig26} J_{\nu -1/2}(k) - \sign(\lambda) \ue^\theta J_{\nu
+1/2 }(k) & = 0 \quad
\text{for the} + \text{case}, \\
\label{eig27} J_{\nu -1/2}(k) + \sign(\lambda) \ue^{-\theta}
J_{\nu +1/2}(k) & = 0 \quad \text{for the} - \text{case}.
\end{align}
The degeneracy of the eigenvalues can be determined by Weyl's
dimension formula. For each $\ell$ in $D=2d$ dimensions and for
both, \refs{eig26} and \refs{eig27}, the degeneracy is given by
\begin{align}
\label{eig28} d_\ell(D) = \frac{d_s}{2} \begin{pmatrix} D+\ell-2 \\
\ell
\end{pmatrix},
\end{align}
with $d_s=2^{d}$ the dimension of the spinor space.
\section{The eta function as an integral for arbitrary $D$}
\label{eta} Our starting point for the evaluation of the spectral
asymmetry will be a contour integral representation of the eta
function, which we derive first; for the general strategy see,
e.g., \cite{BSW,kkbuch}. In order to clearly distinguish positive
and negative eigenvalues, let us write down equations
(\ref{eig26}) and (\ref{eig27}) for $\lambda
> 0$ and $\lambda<0$.

For $\lambda>0$ we have \beq \label{eta15} \nn J_{\nu -1/2
}(k)-\ue^{\Th}
J_{\nu +1/2}(k)&=&0,\\
J_{\nu -1/2}(k)+\ue^{-\Th} J_{\nu +1/2}( k)&=&0,\qquad \ell
=0,...,\infty \,, \eeq while for $\lambda<0$ \beq \label{eta16}
\nn J_{\nu -1/2}(k)+\ue^{\Th}
J_{\nu +1/2 }(k)&=&0,\\
J_{\nu -1/2}(k)-\ue^{-\Th} J_{\nu +1/2 }(k)&=&0,\qquad
\ell=0,...,\infty \,. \eeq Hence, the $\eta$-function \beq \eta
(s,\ui \pslsh) = \sum_\lambda (\mbox{sgn} \lambda) |\lambda| ^{-s}
\nn\eeq is given by the following contour integral in the complex
plane, \beq \label{eta17} \eta(s, \ui \pslsh)=\sum_{\ell
=0}^{\infty} d_\ell (D) \frac 1 {2\pi i} \int\limits_{\Gamma} \ud
z\,z^{-s} \frac{\ud}{\ud z}\log{\left(\frac{J_{\nu -1/2
}(zR)-\ue^{\Th} J_{\nu +1/2}(zR)} {J_{\nu -1/2}(zR)+\ue^{\Th}
J_{\nu +1/2}(zR)}\right)}-(\Th \rightarrow -\Th)\,,\nn \eeq where
the contour $\Gamma$ encloses the positive real axis
counterclockwise. The notation above means, that we have to
subtract the same expression, with $\theta$ replaced by $-\theta$.

We deform the path of integration such that we integrate along the
imaginary axis.
After using the definition of the modified Bessel functions and
the elementary relation \cite{grad} $$ \arctan x = \frac 1 {2\ui}
\ln \frac{ 1+\ui x}{1-\ui x} ,$$ we find \beq \label{eta19} \nn
\eta(s,\ui \pslsh )=\frac 2 {\pi}\cos{\left(\frac{\pi
s}{2}\right)} \sum_{\ell =0}^{\infty} d_\ell (D)
\int\limits_{0}^{\infty} \ud t\,t^{-s} \frac{\ud}{\ud
t}\left[\arctan Q_\nu (\theta ,tR) - \arctan Q_\nu (-\theta
,tR)\right] ,\nn\eeq where the notation
$$Q_\nu (\theta , x) = \ue ^\theta \,\,\frac{ I_{\nu +1/2} (x)}{I_{\nu
-1/2} (x)} $$ has been used.

In order to evaluate $\eta(0,\ui \di)$, which is by definition the
spectral asymmetry, we will use in the forthcoming sections the
shifted Debye expansion of Bessel functions summarized in Appendix
\ref{Debye}. We change the variable in the integral according to
$t= \frac{\nu u}{R}$. We observe that all terms combine nicely if
\cite{grad}
$$\arctan x - \arctan y = \frac \pi 2 -\arctan \frac {1+x y}{x-y} $$
is used. With
$$P_\nu (\theta, x) = \frac 1 {2\sinh \theta} \left(\frac{I_{\nu +1/2} (x)}  {I_{\nu
-1/2} (x)}+\frac{I_{\nu-1/2} (x)}{I_{\nu+1/2} (x)}\right),$$ we
obtain the simple looking result \beq \label{eta20} \eta(s,\ui
\pslsh  )&=&-\frac 2 {\pi } R^s \cos{\left(\frac{\pi
s}{2}\right)}\sum_{\ell=0}^{\infty} d_\ell (D) \nu^{-s}
\int\limits_{0}^{\infty}\ud u\,{ u}^{-s} \frac{\ud} {\ud u}\arctan
P_\nu (\theta, u\nu ). \label{etaclosed}\eeq Using the notation
$$b=\frac{\sqrt{1+u^2}}{u\sinh\theta}, \quad \quad t=\frac 1
{\sqrt{1+u^2}},$$ as $\nu\to\infty$, equations (\ref{I}) and
(\ref{Debye1}) in Appendix \ref{Debye} show that \beq P_\nu
(\theta, \nu u) = b (1+\delta ) \nn\eeq where \beq \delta = \frac
1 \nu \frac 1 2 t^3 + \frac 1 {\nu^2} \frac 5 8 t^4 (1-t^2) +
\frac 1 {\nu^3} t^5 \left( 1 - \frac{23} 8 t^2 + \frac {15} 8
t^4\right) + {\cal O} \left( \nu^{-4}\right) .\nn \eeq Using this
in \beq \arctan \left( b [ 1+\delta]\right) = \arctan b + \delta
\,\,\frac b {1+b^2} - \delta ^2\,\, \frac{b^3}{1+b^2} + \delta ^3
\,\,\frac {b^3} 3 \,\,\frac{ 3b^2 -1}{ (1+b^2)^3 } + {\cal O}
(\delta ^4) ,\nn\eeq the leading three orders of the
$\nu\to\infty$ expansion are obtained. Introducing $Z = 2+u^2 +
u^2 \cosh (2\theta)$, where $Z^{-j}$ naturally occurs as a factor
multiplying $\delta^j$, we find explicitly \beq \arctan P_\nu
(\theta , \nu u) &=& \arctan \left(\frac{
\sqrt{1+u^2}}{2\sinh\theta} \right) \nn\\
&& \hspace{-3.0cm}+\frac{2u\sinh\theta } Z \left\{ \frac 1 \nu
\,\, \frac 1 2 t^2 + \frac 1 {\nu ^2} \,\, \left(\frac 5 8 t^3 -
\frac 5 8 t^5\right) + \frac 1 {\nu^3} \left( t^4 - \frac{23} 8
t^6 + \frac {15} 8
t^8\right) \right\} \nn\\
&& \hspace{-3.0cm} - \frac{4u\sinh \theta} {Z^2} \left\{ \frac 1
{\nu^2} \frac 1 4 t^3 + \frac 1 {\nu^3} \left( \frac 5 8 t^4 -
\frac 5 8 t^6\right) \right\} \nn\\
& &\hspace{-3.0cm} + \frac 1 {\nu^3} \frac {u \sinh \theta} {Z^3}
\left\{ t^4 - \frac 1 3 t^6 u^2 \sinh ^2 \theta \right\}+{\cal O}
\left( \frac 1 {\nu^4}\right) .\label{asympexp} \eeq These
asymptotic contributions are all we will need for the evaluation
of $\eta (0,\ui \pslsh )$ in two and four dimensions. More
general, in $D$ dimensions, $D-1$ asymptotic orders would be
needed.

Note that using (\ref{asympexp}) in (\ref{etaclosed}), each order
of $1/\nu$ leads to the appearance of zeta functions of the Barnes
type \cite{barnes1,barnes2,dowk}, \beq \zeta_{{\cal B}} (s,a) =
\sum_{\ell =0}^\infty {D+\ell -2 \choose \ell} (\ell +a)^{-s}
.\nn\eeq In detail we have \beq \label{eta24} \sum_{\ell
=0}^{\infty}d_\ell(D)\nu^{-s}= \frac 12 d_s \zeta_{{\cal B}}
\left(s,\frac{D-1}{2}\right).\eeq This fact is characteristic for
spectral problems on balls.

All resulting $u$-integrations can be done in terms of
hypergeometric functions, \beq \int\limits_0^\infty du u^{-s}
(1+u^2)^{-\alpha} \left[ 1 + \frac {u^2}{2 (1+u^2)} (\cosh
(2\theta ) -1 )\right] ^{-\beta} &=& \nn\\
& & \hspace{-6.0cm}\frac{ \Gamma \left( \alpha + \frac{ s-1} 2
\right) \Gamma \left( \frac {1-s} 2 \right)} {2\Gamma (\alpha )}
\,\, _2 F _1 \left( \beta , \frac{1-s} 2 ; \alpha ; -\sinh ^2
\theta \right). \label{relint}\eeq This allows us to express all
terms resulting from the asymptotic expansions (\ref{asympexp}) in
the compact form \beq A_{i,k,l,j} (s) &=& -\frac {s d_s} {2^{j+1}
\pi } \cos \left( \frac {\pi s} 2 \right) \frac{ \Gamma \left(
j+\frac{ k+s-l} 2 \right) \Gamma \left( \frac{ l-s} 2 \right) } {
\Gamma \left( j + \frac k 2 \right) } \zeta _{{\cal B} } \left(
s+i ; \frac{ D-1} 2
\right) \times \nn\\
& & \quad \quad _2 F _1 \left( j , \frac{ l-s} 2 ; j + \frac k 2 ;
-\sinh ^2 \theta \right) .\label{aijkl} \eeq We next use these
results to evaluate the asymmetry on balls in two and four
dimensions.
\section{$D=2$: The asymmetry on a disk}
\label{section3} In two dimensions the degeneracy is $d_\ell (2)
=1$ and we have $\nu = \ell+1/2$, such that from (\ref{etaclosed})
we get \beq \eta (s, \ui \pslsh ) = - \frac 2 \pi R^s \cos \left(
\frac {\pi s} 2 \right) \sum_{\ell =0}^\infty \nu^{-s}
\int\limits_0^\infty du u^{-s} \frac d {du} \arctan P_\nu (\theta
, u\nu ) .\label{disk1}\eeq Here, the relevant Barnes boundary
zeta function reduces to a Hurwitz zeta function, i.e.,
$\zeta_{{\cal B}}(s,\frac12)$=$\zeta _H (s,1/2)$. In this case, in
order to obtain the analytic extension to $s=0$, it is enough to
consider only two terms in the Debye expansion. In order to see
this we integrate (\ref{disk1}) by parts and have for $0<\Re
(s)<1$ \beq \label{disk2} \nn \eta(s, \ui \pslsh )=- \frac{2s} \pi
R^s \cos{\left(\frac{\pi s}{2}\right)}\sum_{\ell
=0}^{\infty}{\nu}^{-s} \int\limits_{0}^{\infty}\,\, du \,\,
u^{-s-1} \arctan P_\nu (\theta , u\nu ). \eeq The $s$-factor in
the numerator can be cancelled by singularities coming from
divergencies in the integral (this is the case for the
$\nu$-independent term in the Debye expansion) or from the pole in
the successive Hurwitz zeta functions (this is the case for the
term of order $\nu^{-1}$ in the Debye expansion).

The leading two asymptotic terms contributing to $\eta (0, \ui\di
)$ are \beq \eta (s, \ui \pslsh) = \sinh\theta \left( A_{0,1,1,1}
(s) + A_{1,2,1,1} (s) +...\right) \nn\eeq Because $\zeta _H
(0,1/2) =0$ we have $A_{0,1,1,1} (0)=0$ and this shows \beq \eta
(0, \ui\di) &=& - \frac 1 4 \sinh \theta \,\, _2 F_1 \left( 1 ,
\frac
1 2 , 2 , -\sinh^2 \theta \right) \nn\\
&=&  -\frac 1 2\,\, \frac{\sinh \theta }{1+ \cosh \theta} = -\frac
1 2 \tanh \left( \frac \theta 2 \right)\,. \label{etain2}\eeq Note
that the result is invariant under the transformation $\theta \to
\theta + 2\pi i$.

The case $D=2$ was treated before, in reference \cite{BSW}, as an
example of a non-product manifold. Unfortunately, the term coming
from the pole in the Hurwitz zeta function was missing in that
calculation, which led to erroneous conclusions about relations
between results on product and non-product manifolds stated in the
same reference.
\section{Asymmetry in $D=4$}
\label{four} In the four-dimensional case we have $\nu = \ell+3/2$
and the degeneracy reads $d_\ell (4) = \nu^2 -1/4$. Our immediate
concern is the evaluation of the $\eta$-function at the value
$s=0$ in dimension $D=4$. Arguing as below (\ref{disk1}), terms up
to the order $1/\nu^3$ need to be considered. Further
simplifications occur since (see, e.g., \cite{cmpdk,kkbuch}) \beq
\zeta _{{\cal B}} \left( 0 ; \frac 3 2 \right) = 0, \quad \quad
\mbox{Res } \zeta _{{\cal B}} \left(s=2; \frac 3 2 \right) =0
\,.\nn\eeq As a consequence we find that \beq A_{0,k,l,j} (0) = 0,
\quad \quad A_{2,k,l,j} (0) =0 \nn\eeq for all relevant values of
$k,l,j$. Then, the contributions to the eta invariant read
\beq\eta (0) &=& \sum_{i,k,l,j} C_{i,k,l,j} A_{i,k,l,j} (0)
\label{etaexp} \eeq where $i=1,3$ contributes and the
non-vanishing numerical multipliers are found from
(\ref{asympexp}); in detail, the results are \beq C_{1,2,1,1} =
\sinh \theta & & C_{3,4,1,1} = 2 \sinh \theta \quad C_{3,6,1,1} =
-\frac{23} 4 \sinh \theta \quad C_{3,8,1,1} = \frac{15} 4 \sinh
\theta \nn\\
C_{3,4,1,2} = -\frac 5 2 \sinh \theta & & C_{3,6,1,2} = \frac 5 2
\sinh \theta \quad C_{3,4,1,3} = \sinh \theta \quad C_{3,6,3,3} =
- \frac 1 3 \sinh ^3 \theta .\nonumber \eeq For all relevant
values of $i,k,l,j$, the hypergeometric functions turn out to be
hyperbolic functions. Using (see, e.g., \cite{cmpdk,kkbuch}) \beq
\mbox{Res } \zeta _{{\cal B}} \left( s=3 ; \frac 3 2 \right) =
\frac 1 2 , \quad \mbox{Res } \zeta _{{\cal B}} \left( s=1; \frac
3 2 \right) = -\frac 1 8 ,\nn\eeq and adding up all contributions,
we find \beq \eta (0, \ui\di ) = \frac 1 {6144} \frac{\tanh \left(
\frac \theta 2 \right) }{\cosh ^6 \left( \frac \theta 2 \right) }
\left( 259 + 344 \cosh \theta + 161 \cosh (2\theta ) + 16 \cosh
(3\theta ) \right) . \label{finans}\eeq
\section{Determination of the leading coefficients in heat traces}
\label{coefficients}

In this section, we use the special case of the ball just
presented, together with similar results for the cylinder and
functorial techniques, to find some results about the coefficients
$a_n^\eta (F,P,\Pi_- )$ in the expansion
\cite{bene03-36-11533,gilk95b,seel} \beq \Tr \left( F P e^{-t P^2}
\right) \sim \sum_{n=0}^{\infty} a_n^\eta (F,P, \Pi_- )
t^{\frac{n-D-1} 2} ,\nn\eeq with $P$ an operator of Dirac type
decomposed as $$P= \ui\gamma_j \nabla_j + \psi.$$ Here $\nabla$
denotes a connection on a vector bundle $V$ over a compact
$D$-dimensional Riemannian manifold $M$ with smooth boundary
$\partial M$ such that $\nabla \gamma =0$, furthermore $\psi$ and
$F$ are smooth endomorphisms of $V$.

The functorial techniques will consist of conformal
transformations and of relations between the above expansion and
the well-known expansion for the heat-kernel
\cite{bene03-36-11533,gilk95b,seel} \beq \Tr \left( F e^{-t
P^2}\right) \sim \sum_{n=0}^\infty a_n^\zeta (F,P^2,\Pi_- )
t^{(n-D)/2} . \nn\eeq To write down the geometrical structure of
the coefficients $a_n^\eta$ let $L_{ab}$ be the second fundamental
form, $a,b=1,...,D-1$, and let $_{;D}$ denote the derivative with
respect to the exterior normal.
\begin{lemma}\label{lemx.1}
Let $f$ be scalar, and $F=f\cdot \mbox{Id}_V$. There exist
universal constants $d_i (\theta ,D)$ such that \beq a_0^\eta
(F,P,\Pi_- ) &=& 0,\nn\\
a_1^\eta (F,P,\Pi_- ) &=& (4\pi)^{-D/ 2} \left\{(1-D)
\int\limits_M f \Tr (\psi ) dx + \int\limits_{\partial M} \don f
\Tr (\mbox{Id}) dy
\right\} ,\nn\\
a_2 ^\eta (F,P,\Pi_- ) &=& (4\pi )^{-(D-1)/2}
\int\limits_{\partial M} \Tr \left( \dtw L_{aa} f + \dth f \psi
\gt ( {\rm i} S ) \right.\nn\\
& & \left. + \dfo f \psi ({\rm i} S ) + \dfi f \psi \gt + d_6
(\theta , D) f \psi + d_7 (\theta ,D) f_{;D}\mbox{Id}_V
\right)dy\nn\eeq
\end{lemma}
{\it Proof:} This follows from the theory of invariants as
described e.g. in \cite{gilk95b}.$\quad\Box$

We have used the invariant $\ui S$, with $S$ the projection of
$\gamma^A$ onto the outward unit normal vector, such that the
numerical multipliers remain real as is commonly chosen. We next
determine the universal multipliers $d_i (\theta , D)$,
$i=1,...,7$. We first exploit known special cases.
\begin{lemma}\label{lemx.2}
We have \beq \quad\quad d_1 (\theta ,D) &=& \frac 1 2 (D-1)\sinh
\theta
 \,\, _2F_1 \left( 1,1-\frac D 2 ; \frac 3 2 ; -\sinh ^2 \theta
 \right), \nn\\
 \dtw &=& -\frac 1 {16} (D-2) \sinh \theta \,\, _2F_1 \left( 1 , \frac{3-D} 2 ; 2 ;
 -\sinh ^2 \theta \right).\nn\eeq
\end{lemma}
{\it Proof:} This follows from the calculations on the ball with
$f=1$ and $\psi =0$, which is the situation considered in the
previous sections. For convenience, we also put the radius of the
ball $R=1$. We will use the relation \cite{gilk95b,seel} \beq
a_n^\eta (1,P, \Pi_- )= \frac 1 2 \Gamma \left( \frac{ D-n+1} 2
\right) \mbox{Res } \eta (D-n , P ) \label{etazeta}\eeq and,
therefore, we can use the previously analyzed $\eta$-function
$\eta (s,\ui\di )$.

Instead of looking at $s=0$, the information
relevant for Lemma \ref{lemx.1} is found by considering $\eta
(s,\ui\di )$ in eq. (\ref{etaclosed}) about the points $s=D-1$ and
$s=D-2$. Arguing as before, we need to consider the leading two
asymptotic contributions in (\ref{asympexp}) only and find, modulo
terms that do not contribute to the relevant residues \beq \eta
(s,\ui\di ) &\sim & - \frac 1 {\sqrt \pi} d_s \frac{\Gamma \left(
1 + \frac s 2 \right)}{\Gamma \left( \frac {1+s} 2 \right) } \sinh
\theta \,\, _2 F_1 \left( 1 , \frac{1-s} 2 ; \frac 3 2 ; -\sinh ^2
\theta \right) \zeta _{\cal
B} \left( s ; \frac{ D-1} 2 \right)  \nn\\
 & & +\frac 1 8 d_s \,\, s (s+1) \sinh \theta \,\, _2 F _1 \left(
1 , \frac{ 1-s} 2 ; 2 ; -\sinh ^2 \theta \right) \zeta _{\cal B}
\left( s+1 ; \frac{ D-1} 2 \right) .\nn\eeq The residues of $\eta
(s,\ui\di )$ at $s=(D-1)$ and $s=(D-2)$ follow immediately from
the residues of the Barnes zeta function, see e.g.
\cite{kirs96-54-4188}, \beq \mbox{Res } \zeta _{\cal B} \left( D-1
; \frac{D-1} 2 \right) &=&
\frac 1 {(D-2)!} = \frac 1 {\Gamma (D-1)} ,\nn\\
\mbox{Res } \zeta _{\cal B} \left( D-2 ; \frac{D-1} 2 \right) &=&
0.\nn\eeq Using equation (\ref{etazeta}), and the fact that the
volume of the $(D-1)$-dimensional unit sphere is
$2\pi^{D/2}/\Gamma(D/2)$, we conclude the proof of Lemma
\ref{lemx.2}. $\quad\Box$

We next apply conformal variations of Dirac type operators.
\begin{lemma}\label{lemx.3}
We have $$d_7 (\theta ,D) = \frac{D-1}{D-2} \,\,\dtw .$$
\end{lemma}
{\it Proof:} Let $f$ be a smooth function with $f|_{\partial M}
=0$. Define $g_{\mu\nu}(\epsilon ) := e^{2\epsilon f} g_{\mu\nu}
(0)$ and $P(\epsilon ) := e^{-\epsilon f} P$. Let $\nabla$ be the
standard spinor connection. We expand $P = \ui \gamma^\nu
\nabla_{\partial_\nu} + \psi$ with respect to a local coordinate
system $x=(x_1,...,x_D)$ and use the metric to lower indices and
define $\gamma_\nu$. Then the connection transforms like$$\nabla
(\epsilon ) _{\partial_\mu } := \nabla_{\partial _\mu} + \frac 1 2
\epsilon (- f_{;\nu} \gamma^\nu \gamma_\mu + f_{;\mu}).$$
Furthermore,
$$\psi (\epsilon ) = e^{-\epsilon f} \left( \psi - \frac 1 2
\epsilon (D-1) f_{;\nu} (\ui \gamma^\nu ) \right).$$ Note that the
boundary condition remains unchanged under conformal variation.
Proceeding as for the heat kernel coefficients, one shows that the
eta invariant coefficients satisfy the equation
\cite{gilk05-38-8103}\beq \left. \frac d {d\epsilon}
\right|_{\epsilon =0} a_n ^\eta\left( 1, P (\epsilon ) , \Pi_-
\right) = (D-n) a_n ^\eta\left( f , P , \Pi_- \right).
\label{conf1}\eeq To study the numerical multiplier $d_7 (\theta
,D)$ we need the variation of \beq \left. \frac d {d\epsilon}
\right|_{\epsilon =0}L_{aa} &=& - f L_{aa} + (D-1) f_{;D} .\nn\eeq
Applying equation (\ref{conf1}) shows the assertion. $\quad\Box$

In order to determine the numerical multipliers $\dth , \dfo ,
\dfi$ and $\dsi$ we relate the eta invariant to the zeta
invariant. We will then evaluate the zeta invariant on the
$D$-dimensional cylinder for the case of an arbitrary endomorphism
valued $F$.

The result we are going to use is the following:
\begin{lemma}\label{lemx.4}
Let $F \in C^\infty (\mbox{End} (V))$ and let $P (\epsilon ):= P +
\epsilon F$. We then have \beq
\partial _\epsilon a_n ^\eta (1, P (\epsilon ), \Pi_-
) = (n-D) a_{n-1} ^\zeta (F , P(\epsilon ), \Pi_- ).\nn\eeq
\end{lemma}
{\it Proof:} The proof is insensitive to the boundary conditions
imposed and parallels the proof in
\cite{bran92-108-47,gilk05-38-8103}. $\quad\Box$
\begin{remark}\label{remx.5} The very useful property of this result is that the
$a_n^\eta$ coefficient for the eta invariant is related to the
coefficient $a_{n-1}^\zeta$ for  the zeta invariant, which will
have a significantly simpler structure. \end{remark} In order to
apply Lemma \ref{lemx.4} to the coefficient $a_2^\eta$ we need the
general form of the $a_1^\zeta$ coefficient.
\begin{lemma}\label{lemx.6} Let $F\in C^\infty (\mbox{End} (V))$.
There exist universal constants $f_i (\theta , D)$ such that \beq
& & a_1 ^\zeta (F,P^2,\Pi_- ) = \nn\\
& &\quad (4\pi)^{-(D-1)/2} \int\limits_{\partial M} \Tr \left\{
f_3 (\theta ,D) F\gt ({\rm i} S) + f_4 (\theta ,D) F({\rm i} S ) +
f_5 (\theta ,D)  F\gt + f_6 (\theta ,D) F\right\} .\nn\eeq
\end{lemma}
{\it Proof:} This follows immediately from the theory of
invariants taking into account that $F$ is in general a
matrix-valued endomorphism. $\quad\Box\\$
\begin{remark}\label{remx.7} Lemma \ref{lemx.4} relates the
universal constant $d_j (\theta ,D)$ with $f_j (\theta ,D)$,
$j=3,...,6$. In detail we have \beq d_j (\theta ,D) = -(D-2) f_j
(\theta ,D) , \quad j=3,...,6. \nn\eeq Finding the $f_j (\theta
,D)$ is easier, because they follow from the case with $\psi =0$.
They can be evaluated from the cylinder where the heat kernel is
known locally \cite{bene03-36-11533} and its coefficients can be
evaluated for an arbitrary endomorphism $F$.
\end{remark}
In order to summarize the results of \cite{bene03-36-11533} we
need to introduce the relevant notation. Let $M=\reals_+ \times N$
be an even dimensional cylinder equipped with the metric $ds^2 =
dx_D^2 + ds_N^2$, where $x_D$ is the coordinate in $\reals_+$ and
plays the role of the normal coordinate, and $ds_N^2$ is the
metric of the closed boundary $N$. The coordinates on $N$ are
denoted by $y=(y_1,y_2,...,y_{D-1})$. To write down the heat
kernel on $M$ for $P^2= (\ui\gamma_j \nabla_j)^2$ with boundary
condition $\Pi_-$, we call $\phi_\omega (y)$ the eigenspinors of
the operator $B=\gt S \gamma_a \nabla _a$, corresponding to the
eigenvalue $\omega$, normalized so that \beq \sum_{\omega}
\phi_{\omega}^{\star}(y) \phi_{\omega}(y^{'})=
\delta^{D-1}(y-y^{'}), \nn\eeq with $\delta^{D-1}$ the Dirac delta
function, and \beq \int\limits_{N}\,\,dy\,\,
\phi_{\omega}^{\star}(y) \phi_{\omega}(y)=1\,.\nn\eeq Finally we
need $x=(y,x_D)$, $\xi = x_D - x_D '$, $\eta = x_D + x_D'$,
$u_\omega (\eta ,t) = \frac{\eta}{\sqrt {4t}} + \sqrt t \omega
\tanh \theta$, and the complementary error function
\[\mbox{erfc}(x)=\frac{2}{\sqrt{\pi}}\int_{x}^{\infty}d\xi
e^{-\xi^2}\, .\] We then have the result.
\begin{lemma}\label{lemx8} From the calculation on the cylinder $M=\reals_+ \times N$
we obtain the following values for the multipliers $f_i(\theta
,D)$,
$i=3,4,5,6$: \beq f_3 (\theta ,D) &=& \frac 1 4 \cosh ^{D-2} \theta , \nn\\
 f_4 (\theta , D)&=& 0 , \nn\\
 f_5 (\theta , D)&=& \frac 1 4 \cosh ^{D-2} \theta \sinh \theta , \nn\\
 f_6 (\theta , D)&=& \frac 1 4 \left( \cosh ^{D-1} \theta -1 \right) .\nn\eeq
 \end{lemma}
{\it Proof:} In \cite{bene03-36-11533} we have shown that the
local heat-kernel on the cylinder reads \beq K(x,x';t) &=&
\frac{1}{\sqrt{4\pi t}}\sum_{\omega}\phi_{\omega}^{\star}(y^{'})
\phi_{\omega}(y) e^{-\omega^2
t}\left\{\left(e^{\frac{-\xi^2}{4t}}-e^{\frac{-\eta^2}{4t}}\right)\mathbf{1}\right.\nn\\
& & \left.+\frac{2\pip \pipl}{\cosh^2( \theta )}\left[1-\sqrt{(\pi
t)}\omega \tanh{\Th}e^{u_{\omega}(\eta,t)^2}
\mbox{erfc}(u_{\omega}(\eta,t))\right]e^{\frac{-\eta^2}{4t}}\right\}
, \label{hkcyl}\eeq with $\Pi_+ = (1/2) (1+\ui e^{\theta \gt} \gt
S)$. (Note that \cite{bene03-36-11533} and the present article use
different conventions. Here we use the exterior normal contrary to
the interior normal there. Furthermore, the $\gamma_*$ here is
minus $\tilde \gamma$ there. As a result, in the solution formula
from \cite{bene03-36-11533} we have to replace $\theta$ by
$-\theta$ in order to find a solution for the problem considered
here.) The first term is the heat-kernel of the manifold $\reals
\times N$, a manifold that has no boundary. Therefore we do not
consider this term further as it provides no relevant information
for Lemma \ref{lemx.1}.

It is natural to introduce the heat kernel $K_B (y,y'; t)$ of the
operator $B^2$, \beq K_B (y,y'; t) = \sum_\omega \phi _\omega ^*
(y') \phi _\omega (y) e^{-\omega ^2 t } ; \nn\eeq furthermore, to
make the single steps easier to follow we use the splitting \beq
K_1 (x,x'; t) &=& -\frac{1}{\sqrt{4\pi
t}}\sum_{\omega}\phi_{\omega}^{\star}(y') \phi_{\omega}(y)
e^{-\omega^2 t}e^{\frac{-\eta^2}{4t}}, \nn\\
K_2 (x,x' ;t) &=& \frac{1}{\sqrt{4\pi
t}}\sum_{\omega}\phi_{\omega}^{\star}(y') \phi_{\omega}(y)
e^{-\omega^2 t}\frac{2\pip
\pipl}{\cosh^2(\Th)}\nn\\
& &\left[1-\sqrt{(\pi t)}\omega \tanh{\Th}e^{u_{\omega}(\eta,t)^2}
\mbox{erfc}(u_{\omega}(\eta,t))\right]e^{\frac{-\eta^2}{4t}} .
\nn\eeq We are interested in the trace $\Tr _{L^2} ( F K
(x,x;t))$. We assume $F=F(y)$ to be independent of the normal
variable $x_D$, such that the $x_D$-integration of the $L^2$-trace
can be done without greater complication.

First, it is straightforward to see that \beq \int\limits_0^\infty
dx_D F(y) K (x,x;t) = - \frac 1 4 F(y) K_B (y,y;t).
\label{mariel1}\eeq The representation for $K_2 (x,x;t)$ can be
conveniently rewritten as to perform the $x_D$-integration. We
have \beq K_2 (x,x;t) &=&
  -\frac 1 {2 \cosh ^2 \theta} \sum_\omega \phi^*_\omega (y) \phi_\omega
(y) e^{-\omega ^2 t} \Pi_+ \Pi_+^* \nn\\
& & \quad \quad \frac \partial {\partial x_D} \left[ e^{ -\frac
{x_D^2} t + u_\omega ^2 (2 x_D , t) } \mbox{erfc} \left( u_\omega
(2x_D , t) \right) \right] ,\nn\eeq where we used the relation
\[ -\frac12 \frac{\partial}{\partial x_D}
\left[e^{-x_D^2/t+u_{\omega}^2(2x_D,t)} \mbox{erfc}
(u_{\omega}(2x_D,t))\right] =\]\beq
e^{-x_D^2/t}\left[\frac{1}{\sqrt{\pi t}}-\omega \tanh\theta\,
e^{u_{\omega}^2(2x_D,t)}
\mbox{erfc}(u_{\omega}(2x_D,t))\right]\,.\nn\eeq Therefore, \beq
\int\limits_0^\infty dx_D F K_2 (x,x;t) &=& \frac 1 {2 \cosh ^2
\theta} \sum_\omega \phi^*_\omega (y) \phi_\omega (y) e^{-\omega
^2 t}\Pi_+ \Pi_+^* \nonumber\\
& &\hspace{3.0cm} e^{t\omega^2 \tanh ^2 \theta } \mbox{erfc}
\left( \sqrt t \omega \tanh \theta \right) .\nn\eeq We need to
collect the contributions to the coefficient $a_1^\zeta (F,P^2,
\Pi_- )$. The first term, equation (\ref{mariel1}), can be
described by the heat-kernel of the boundary and we find the
relevant contribution to be $-(1/4) a_0 ^\zeta (F, B^2)$. In order
to find the contribution of $K_2 (x,x;t)$ is considerably harder;
we found it most convenient to relate the heat-kernel coefficients
to the zeta function, \beq \zeta (s;F,P^2 , \Pi_- ) = \frac 1
{\Gamma (s)} \int\limits_0 ^\infty dt \,\, t^{s-1} \Tr _{L^2}
\left( F e ^{-t P^2}\right),\nn\eeq in the standard fashion \beq
\mbox{Res }\zeta \left( z ; F,P^2, \Pi_- \right) =
\frac{a_{D-2z}^\zeta (F,P^2,\Pi_- )}{\Gamma (z)} ,\eeq for
$z=D/2,(D-1)/2,...,1/2,-(2n+1)/2,n\in\nats$.

For $K_2 (x,x;t)$ the related zeta function contribution is \beq
\zeta _2 (s,F,P^2 , \Pi_-) &=& \frac 1 {2 \cosh ^2 \theta\Gamma
(s)} \Tr \left\{F \sum_\omega\phi^*_\omega (y) \phi_\omega
(y)\Pi_+
\Pi_+^* \times \right.\nn\\
& & \left.\int\limits_0 ^\infty dt \,\, t^{s-1} e^{-\frac{\omega
^2 t}{\cosh ^2 \theta}}\left( 1 + \mbox{erf} ( -\sqrt t \omega
\tanh \theta ) \right) \right\}.\nn\eeq The integral can be
performed in terms of a hypergeometric function to read \beq \zeta
_2 (s,F,P^2 , \Pi_-) &=& \frac 1 {2 \cosh ^2 \theta\Gamma (s)} \Tr
\left\{F \sum_\omega\phi^*_\omega (y) \phi_\omega (y)\Pi_+
\Pi_+^* \times \right.\nn\\
& &\left.\hspace{-4.50cm}\frac{\cosh^{2s} \theta }{| \omega|^{2s}}
\left[ \Gamma (s) - \frac 2 {\sqrt \pi} \Gamma \left( s + \frac 1
2 \right) \sinh \theta \,\,\mbox{sgn} (\omega) \,\, _2 F _1 \left(
\frac 1 2 , s+ \frac 1 2 ; \frac 3 2 ; -\sinh ^2 \theta \right)
\right] \right\} .\nn\eeq In terms of the boundary spectral
functions this is \beq \zeta _2 (s;F,P^2 , \Pi_-) &=& \frac {\cosh
^{2s-2} \theta } {2 \Gamma (s)} \left\{
\Gamma (s) \zeta \left( s ; \Pi _+ \Pi _+ ^* F, B^2 \right) \right.\label{mariel2}\\
 & & \left.\hspace{-4.0cm} - \frac 2 {\sqrt \pi } \Gamma \left( s + \frac 1 2 \right)
 \sinh \theta \,\, _2 F _1 \left( \frac 1 2 , s+ \frac 1 2 ; \frac
 3 2 ; - \sinh ^2 \theta \right) \eta (2s ; \Pi _+ \Pi _+ ^* F , B
 ) \right\}.\nn\eeq
In order to find the heat-kernel coefficient $a_1^\zeta$, we use
the relation \beq \mbox{Res } \zeta\left( \frac {D-1} 2 ; F , P^2,
\Pi_- \right) = \frac { a_1^\zeta\left(F , P^2, \Pi_-
\right)}{\Gamma \left( \frac{D-1} 2 \right) } .\nn\eeq The eta
invariant in (\ref{mariel2}) does not contribute as $\eta
(D-1;\Pi_+ \Pi _+^* F,B)=0$. Therefore, \beq \mbox{Res } \zeta _2
\left( \frac{D-1} 2 ; F , P^2 , \Pi_- \right) = \frac 1 2 \cosh
^{D-3} \theta \,\,\frac{a_0 ^\zeta \left( \Pi _ + \Pi _ + ^* F ,
B^2 \right)} { \Gamma \left( \frac { D-1} 2 \right) } \nn\eeq and
the contribution to the heat-kernel coefficient $a_1^\zeta (
F,P^2,\Pi_- )$ is \beq \frac 1 2 \cosh^{D-3} \theta \,\,(4\pi )
^{-\frac {D-1} 2 } \int\limits_N dy \,\, \Tr ( \Pi _+ \Pi _+ ^* F
) .\nn\eeq To compare it with the form given in Lemma \ref{lemx.6}
we use \beq \Pi _+ \Pi _+ ^* = \frac 1 2 \cosh \theta \left( \cosh
\theta + \gt \sinh \theta + \gt (\ui S) \right), \nn\eeq providing
the relevant heat-kernel contribution in the form \beq \frac 1 4
\cosh ^{D-2} \theta \,\,(4\pi )^{-\frac {D-1} 2 } \int\limits_N dy
\,\, \Tr \left( \cosh \theta F + \sinh \theta \gt F + \gt (\ui S )
F \right) .\nn\eeq Adding the contribution from $K_1 (x,x;t)$
shows the result. $\quad\Box\\$
\section{Conclusions}
In this article we have determined the asymmetry $\eta (0,\ui
\pslsh )$ of the Dirac operator with chiral bag boundary
conditions given by (\ref{eig5}) on the two-dimensional and
four-dimensional ball, see equations (\ref{etain2}) and
(\ref{finans}). Furthermore, the leading coefficients in the trace
of the smeared kernel corresponding to the eta function were
obtained, see Lemma 6.1, 6.2, 6.3 and 6.8.
\section*{Acknowledgements}
We thank J.D. Lange for very interesting and helpful discussions.
KK acknowledges support by the Baylor University Summer Sabbatical
Program and by the Baylor University Research Committee. EMS was
partially supported by Universidad Nacional de La Plata (Proyecto
11/X381) and Consejo Nacional de Investigaciones Cientificas y
Tecnicas (PIP 6160).

\begin{appendix}
\section{Debye expansion of Bessel functions}
\label{Debye}
Consider the ordinary differential equation \beq
(\partial_z^2 + u
\partial _z + v) \phi (z) =0 .\nn\eeq
For the solution we use the ansatz \beq \phi (z) = \exp \left\{
\int dt\,\, p(t) \right\} \psi (z) .\nn\eeq With the choice
$p=-u/2$, the resulting differential equation for $\psi $ is \beq
\psi '' + q \psi =0, \quad \mbox{where}\quad q = v - \frac {u'} 2
- \frac{u^2} 4 .\nn\eeq It is more convenient to consider \beq
\partial _z \ln \phi = p + S, \quad \mbox{where} \quad S =
\partial _z \ln \psi .\nn\eeq
Contact between $S$ and the solution $\phi$ is made by observing
that \beq \phi (z) = \mbox{const} \,\, \exp \left\{\int dz \,\,
p(z) \right\} \,\, \exp \left\{\int dz \,\, S (z) \right\}
.\nn\eeq The differential equation for $S$ turns out to be \beq S'
= - q - S^2 .\nn\eeq We assume that the differential equation
contains a parameter $\nu$ and that we are interested in the
large-$\nu$ asymptotic of $S$. The particular choice of the
asymptotic expansions below is the relevant case for the
consideration of the asymptotic of Bessel functions. We assume
that as $\nu \to \infty$, the function $q$ has the asymptotic
expansion \beq q = \sum_{i=-2}^\infty \nu^{-i} q_i .\nn\eeq In
that case, the function $S$ can be seen to have the asymptotic
form \beq S=\sum_{i=-1}^\infty \nu ^{-i} S_i .\nn\eeq Using these
asymptotic forms in the differential equation for $S$, the
asymptotic orders are seen to be given by \beq S_{-1} &=& \pm
\sqrt{ - q_{-2} } ,\nn\\
S_0 &=& - \frac{ q_{-1}}{2 S_{-1}} - \frac 1 2 \partial _z \ln
S_{-1} ,\nn\\
S_{i+1} &=& - \frac{ q_i + S_i ' + \sum_{j=0}^i S_j S_{i-j}} { 2
S_{-1}} .\nn\eeq Let us apply these formulas to the differential
equation for the Bessel functions $I_{\nu + \alpha } (\nu z)$ and
$K_{\nu + \alpha } (\nu z)$. The relevant differential equation
reads \beq \phi '' + \frac 1 z \phi ' - \left\{ \nu^2 \left( 1 +
\frac 1 {z^2} \right) + \lambda \frac{ 2 \alpha} {z^2} + \frac{
\alpha ^2} {z^2} \right\} \phi =0 .\nn\eeq Therefore, for the
given example, we find \beq p= - \frac 1 {2z}, \quad q_0 = \frac 1
{z^2} \left( \frac 1 4 - \alpha ^2\right) , \quad q_{-1} = -
\frac{ 2\alpha} {z^2} , \quad q_{-2} = - \left( 1 + \frac 1 {z^2}
\right) .\nn\eeq Linearly independent solutions of the
differential equation are proportional to $I_{\nu +\alpha} (\nu
z)$ and $K_{\nu + \alpha} (\nu z)$. From the known behaviour of
these functions for large arguments \cite{grad}, we can conclude
that $S_{-1} = + \sqrt{-q_{-2}}$ corresponds to $I_{\nu +\alpha}$,
whereas $S_{-1} = - \sqrt{ - q_{-2}}$ corresponds to $K_{\nu
+\alpha}$. For the present occasion we need the asymptotics for
$I_{\nu +\alpha}$ and continue with this case only. Using the
asymptotic orders in $S$, the solution has the asymptotic behavior
\beq \phi (z) \sim \mbox{const} \,\, \exp \left\{ \nu \int dz
S_{-1}\right\} \,\, \exp \left\{ \int dz \left( S_0 - \frac 1
{2z}\right)\right\} \,\, \exp \left\{ \sum_{i=1}^\infty \nu^{-i}
\int dz S_i \right\} . \nn\eeq The constant prefactor is
determined from the known Debye expansion of the Bessel function
$I_\nu (\nu z)$ and it reads
$$\mbox{const} = \frac 1 {\sqrt{2\pi \nu }}.$$ Using the
explicit form of $q_{-2}$ and $q_{-1}$ for the present example,
one obtains therefore the result \beq I_{\nu + \alpha} (\nu z)
\sim \frac 1 {\sqrt{2 \pi \nu}} e^{\nu \beta} t^{1/2} \left(
\frac{ 1-t} {1+t} \right) ^{\alpha /2} \exp \left\{
\sum_{i=1}^\infty \nu^{-i} C_i (z,\alpha)\right\} ,\label{I}\eeq
where $t=1/ \sqrt{1+z^2}$, $\beta = \sqrt{1+z^2} + \ln [ z /
(1+\sqrt{1+z^2})]$,
$$C_i (z,\alpha ) = \int \,\, dz S_i ,$$ and with the understanding
that for $\alpha =0$ the known answers for $I_\nu (\nu z)$ are
reproduced \cite{abra}. The list of the first $C_i (z,\alpha )$
obtained are \beq C_1 (z,\alpha ) &=& \frac{t}{8} -
\frac{{\alpha}^2\,t}{2} -
  \frac{\alpha\,t^2}{2} - \frac{5\,t^3}{24} ,\nn\\
C_2 (z,\alpha ) &=& \frac{t^2}{16} - \frac{{\alpha}^2\,t^2}{4} -
  \frac{13\,\alpha\,t^3}{24} + \frac{{\alpha}^3\,t^3}{6} -
  \frac{3\,t^4}{8} + \frac{{\alpha}^2\,t^4}{2} +
  \frac{5\,\alpha\,t^5}{8} + \frac{5\,t^6}{16},\nn\\
  C_3 (z,\alpha ) &=& \frac{25 t^3}{384} - \frac{13 {\alpha}^2 t^3}{48} +
  \frac{{\alpha}^4 t^3}{24} -\frac{7 {\alpha} t^4}{8}+\frac{{\alpha}^3 t^4}{2}-
\nn \frac{531 t^5}{640} + \frac{7 {\alpha}^2 t^5}{4} -
\frac{{\alpha}^4 t^5}{8} \nn\\&+& \frac{11 {\alpha} t^6}{4} -
\frac{2 {\alpha}^3 t^6}{3} + \frac{221 t^7}{128} - \frac{25
{\alpha}^2 t^7}{16} - \frac{15 {\alpha} t^8}{8} - \frac{1105
t^9}{1152} .\label{Debye1}\eeq Higher orders can be produced as
needed.

\end{appendix}


\end{document}